\documentclass[iop]{emulateapj}
\usepackage{epsfig,amsmath,amsfonts,amssymb,graphicx,subfigure,enumitem}
\shorttitle{Joshi et al.}
\shortauthors{Joshi et al.}

\slugcomment{The Astrophysical Journal}

\begin{document}
\title{PALOMA: A magnetic CV between Polars and Intermediate Polars}
\author{Arti Joshi\altaffilmark{1}}
\author{J.C.Pandey\altaffilmark{1}}
\author{K. P. Singh\altaffilmark{2}}
\author{P.C. Agrawal\altaffilmark{3}}
\affil{\altaffilmark{1}{Aryabhatta Research Institute of Observational Sciences (ARIES), Nainital-263002, India}}
\affil{\altaffilmark{2}{Tata Institute of Fundamental Research, Mumbai-400005, India}}
\affil{\altaffilmark{3}{UM-DAE Centre for Excellence in Basic Sciences, Santacruz (east), Mumbai-400098, India}}

\begin{abstract}
We present analyses of archival X-ray data obtained from the {\it XMM-Newton} satellite and optical photometric data obtained from 1 m class telescopes of ARIES, Nainital of a magnetic cataclysmic variable (MCV) Paloma. Two persistent periods at 156 $\pm$ 1 minutes and 130 $\pm$ 1 minutes are present in the X-ray data, which we interpret as the orbital and spin periods, respectively. These periods are similar to those obtained from the previous as well as new optical photometric observations.  The soft-X-ray excess seen in the X-ray spectrum of Paloma and the averaged X-ray spectra are well fitted by two-temperature plasma models with temperatures of 0.10$_{-0.01}^{+0.02}$ and 13.0$_{-0.5}^{+0.5}$ keV with an Fe K$\alpha$ line and an absorbing column density of 4.6 $\times$ 10\textsuperscript{22} cm \textsuperscript{-2}. This material partially covers 60 $\pm$ 2 \% of the X-ray source. We also present the orbital and spin-phase-resolved spectroscopy of Paloma in the $0.3 - 10.0$ keV energy band and find that the X-ray spectral parameters show orbital and spin-phase dependencies. New results obtained from optical and X-ray studies of Paloma indicate that it belongs to a class of a few magnetic CVs that seem to have the characteristics of both the polars and the intermediate polars.

\end{abstract}


\section{Introduction}
Magnetic cataclysmic variables (MCVs) are evolved, semi-detached interacting binaries containing a magnetic white dwarf (WD), which accretes material from a Roche lobe filling red dwarf star \citep{1995CAS....28.....W}. These stars are characterized by strong X-ray emission, high-excitation optical spectra, very stable X-ray and optical periods in their light curves and optical polarization. MCVs are broadly classified into two categories: polars and intermediate polars (IPs). Polars usually show synchronous or near-synchronous rotation of the magnetized WD with the orbital motion of the binary system and have a magnetic field of $\geq$ 10 MG. The high magnetic field strength in the polars causes the accretion stream flows along the magnetic field lines to its poles and suppresses the formation of the accretion disk leading to a synchronous motion. However, there are some MCVs (e.g., V1432 Aql, BY Cam, CD Ind, and V1500 Cyg) for which the spin period ({\it P}$_{\omega}$) and the orbital period ({\it P}$_{\Omega}$) differ by about 2 \% or less. These systems are thought to be polars and asynchronism is thought to arise due to a recent nova explosion \citep{2004ASPC..315..216N}. The IPs possess a magnetic field of less than 10 MG and the WD rotation is not synchronized with the orbital motion \citep[see][]{2002ASPC..261...92H}. The weaker magnetic field in the IPs allows the material from the companion red dwarf star to form an accretion disk, though the inner part of the disk is disrupted into accretion curtains and columns before dumping the material to the magnetic poles of the WD \citep{1988MNRAS.231..549R}. There are two other systems namely RX J0524+42 (Paloma) and V697 Sco, which are described as nearly synchronous IPs, with {\it P}$_\omega$ = ($0.7 - 0.9$) {\it P}$_\Omega$. Paloma also lies in the period gap and is thought to be currently in the process of attaining synchronism and evolving into a polar \citep{2004ASPC..315..216N, 2007A&A...473..511S}. The orbital period distribution of MCVs shows that the {\it P}$_\Omega$ of polars are in the range of $1.0 - 10.0$ hr, while a majority of the IPs have orbital periods longer than the period gap of $2 - 3$ hrs \citep[]{2010MNRAS.401.2207S}. The period gap is no longer seen to exist for polars \citep{2004ASPC..315..216N, 2010MNRAS.401.2207S}.  

The X-ray luminosities of IPs are greater than those of polars by a factor of $\sim$ 10, attributed mainly to the higher accretion rates \citep[e.g.,][]{1991ApJ...375..600C}. The simplest scenario for X-ray production in MCVs is that the magnetically channeled accretion column impacts the poles of WD, with supersonic velocities leading to the formation of strong shocks. In the post-shock region, the matter cools via thermal bremsstrahlung and cyclotron radiation \citep{1973PThPh..49.1184A, 1994ApJ...426..664W, 1999MNRAS.306..684C} depending upon the magnetic field strength of WD. Therefore, a majority of the IPs are hard-X-ray emitters because the hot plasma of the post-shock region cools on the surface of the WD via thermal bremsstrahlung and the temperature of the post-shock region is $\sim$ $50 - 600$ MK \citep{1995CAS....28.....W, 2001cvs..book.....H}. Some IPs (PQ Gem, V405 Aur, UU Col, and MU Cam) have been recognized to possess a soft-X-ray emission component \citep[see][]{1995A&A...297L..37H, 1996A&A...310L..25B, 2003A&A...406..253S}.  In polars, cyclotron cooling is more efficient \citep{1979ApJ...234L.117L}; therefore, a part of the downward emitted hard X-rays and cyclotron radiation is absorbed by the WD surface and re-emitted in the soft X-rays. 

In this work, we present a detailed analysis of X-ray and optical photometric  observations of Paloma. This object was first characterized by \citet{2007A&A...473..511S} using optical photometry and {\it ROSAT} X-ray observations. They reported an orbital period of 157 minutes and two spin periods of 146 and 136 minutes. With an orbital period in the period gap, \citet{2007A&A...473..511S} suggested that Paloma might be the first bona fide transition object between the DQ Her (IPs) and AM Her (Polars) system with a WD currently in the process of synchronization. The paper is organized as follows. In the next section, we present the observations and data reduction. Section 3 contains an analysis and the results of the X-ray and optical observations, including X-ray and optical timing analyses as well as an X-ray spectral analysis and the result of the phase-resolved X-ray spectroscopy. Finally, we present a discussion and summary in section 4 and 5, respectively.

\section{Observations And Data Reduction}
\subsection{X-Ray Observations}
Paloma was observed with the {\it XMM-Newton} satellite \citep{2001A&A...365L...1J} using European Photon Imaging Camera \citep[EPIC,][]{2001A&A...365L..18S, 2001A&A...365L..27T} and Reflecting Grating Spectrometer \citep[RGS;][]{2001A&A...365L...7D} instruments on 2006 February 22 at 23:23:47 UT for 60 ks in revolution 1137. The EPIC instrument consists of three cameras containing MOS1 and MOS2 \citep{2001A&A...365L..27T} and PN \citep{2001A&A...365L..18S} CCDs, providing extremely sensitive imaging observations over a field of view (fov) of 30 arcmins and an energy range of $0.2 - 10.0$ keV. We used standard {\it XMM-Newton} Science Analysis System (SAS) software package (version$-14.0.0$) for data reduction with updated calibration files. The preliminary processing of raw EPIC Observation Data Files was done using the {\sc epchain} and {\sc emchain} tasks, which allow calibration both in energy and astrometry of the events registered in each CCD chip and combine them in a single data file for MOS and PN detectors. The background contribution is particularly relevant at high energies, therefore, for further analysis, we have selected the energy range between 0.3 and 10.0 keV. Calibrated and concatenated event list files were extracted using SAS tasks {\sc evselect}. Data were corrected to the barycenter of the solar system with the {\sc barycen} task. We have also checked the data for high background proton flares and data that were free from this effect. Subsequently, the  {\sc epatplot} task was used for checking the existence of pile up but it was not observed in the CCD during the observations. The X-ray light curves and spectra of Paloma were extracted from a circular region with a radius 20 arcsecs around the source. The background was chosen from several source-free regions each with a radius of 20 arcsecs on the detectors surrounding the source. Furthermore, the SAS task {\sc eregionanalyse} was also used for the final region selection. Data with single and double-pixel events, i.e., PATTERN = 0$-$4 for PN and PATTERN = 0$-$12 for MOS with FLAG = 0 were used for further analysis. We used the tool {\sc epiclccorr} to correct for good time intervals, dead time, exposure, point-spread function, quantum efficiency, and background  subtraction. The sky region surrounding the target in MOS CCDs was contaminated by events from another region of the sky accumulated during the slew. To solve this problem, we removed the initial one hour of observations because the attitude information allowed us to see when the instrument was on target and it showed that the pointing was stable after removing one hour of observation. We, therefore, created a good time interval file to remove all the events pertaining to the period before the stable pointing. The SAS task {\sc especget} was used to generate the background corrected spectra, which also computes the photon redistribution matrix and the ancillary matrix. The spectra were rebinned for a minimum of 20 counts per bin using grppha. {\sc rgsproc} was run in order to extract the RGS event files and spectra. Further temporal and spectral analyses were done by HEASOFT version 6.17.

\begin{figure}
\centering
\hspace{-0.5 cm}
\includegraphics[width=95mm]{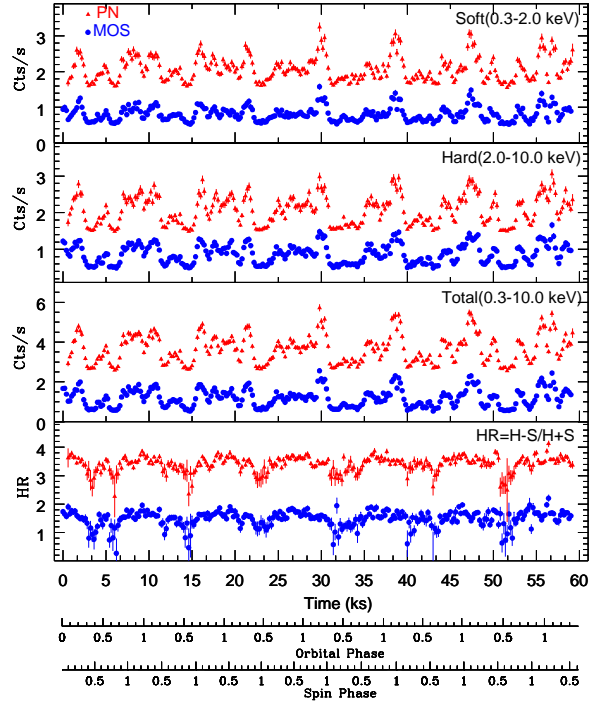}
\caption{X-ray light curves of Paloma at three energy bands named soft, hard, and total. The bottom panel is an HR curve, where HR is defined as (H$-$S)\small/(H$+$S); H and S are count rates in hard and soft energy bands, respectively. In each panel, the upper curve is for PN data while the lower curve is for MOS data. Orbital and spin phases are also shown.}
\label{fig:pnmoslc}
\end{figure}

\begin{figure}
\centering
\hspace{-0.5 cm}
\includegraphics[width=90mm]{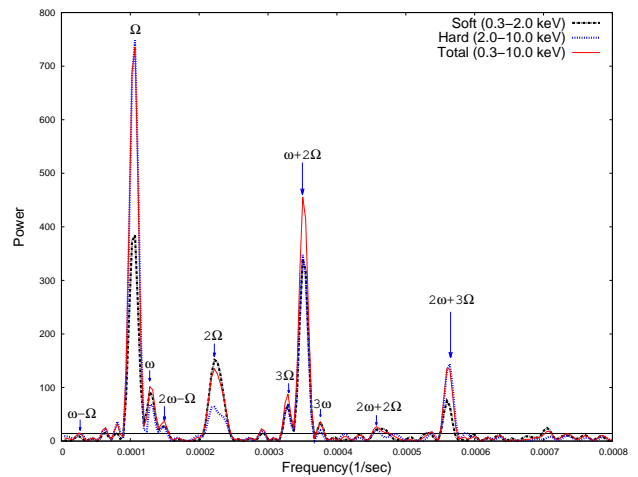}
\caption{Power spectra of Paloma obtained from the EPIC-PN data in the soft, hard, and total energy bands.} 
\label{fig:powspeclc}
\end{figure}
\noindent

\begin{figure}
\centering
\subfigure[]{\includegraphics[width=85mm,angle=0]{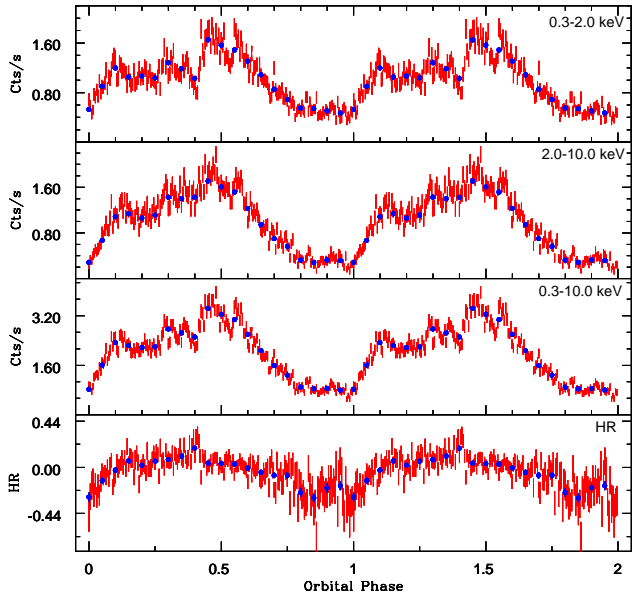}}
\subfigure[]{\includegraphics[width=85mm,angle=0]{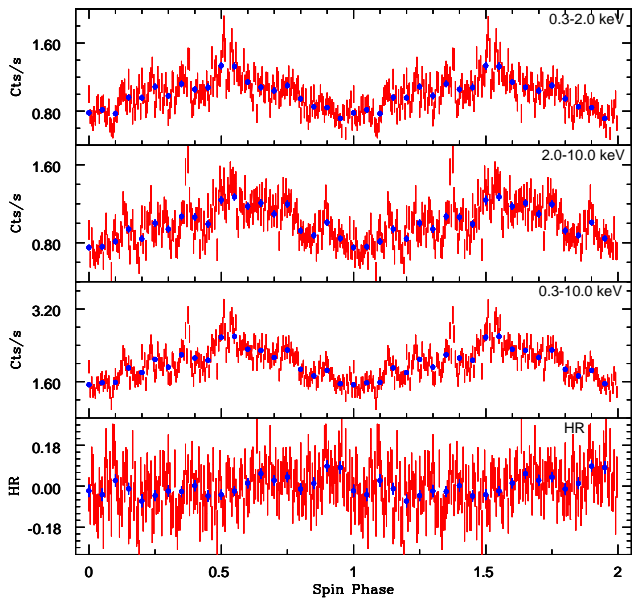}}
\caption{Folded X-ray light curves of Paloma in soft, hard, and total energy bands, where solid dots are folded light curve with phase bin of 0.05. The folded HR curve is also shown in the bottom panels.} 
\label{fig:pnmosfoldlc}
\end{figure}

\subsection{Optical Observations}

R-band photometric observations of Paloma were obtained in 2014 October and November using the 1.04 m Sampurnanand Telescope \citep[ST;][]{1972oams.conf...20S} at Nainital and 1.3 m Devasthal Fast Optical Telescope (DFOT) located at Devasthal, India \citep{2011CSci..101.1020S}. The ST has Ritchey$-$Chretien (RC) optics with a {\it f}\small/13 beam at the Cassegrain focus. The ST is equipped with a 2k $\times$ 2k CCD (read noise = 13.7 $e^{\rm -}$ and gain = 10 $e^{\rm -}$/ADU). Each pixel of the CCD chip has a dimension of 24 $\mu$m$^2$ and it covers an fov of 13\hbox {$^\prime$} $\times$ 13\hbox {$^\prime$}. Observations were carried out in a 2 $\times$ 2 pixel$^2$ binning mode in order to increase the signal-to-noise ratio. The 1.3 m is also RC design with a {\it f}\small/4 beam at the Cassegrain focus. The telescope is equipped with a 2k $\times$ 2k Andor CCD (read noise = 7$e^{\rm -}$ and gain = 2 $e^{\rm -}$/ADU at 1000 kHz readout speed) with a pixel size of 13.5 $\mu$m$^2$. The fov of CCD is $\sim$ 18\hbox {$^\prime$} $\times$ 18\hbox {$^\prime$}. The exposure time per frame was five minutes in both the telescopes. Several bias and twilight sky flat frames were also taken during the observing runs. The pre-processing (i.e., bias subtraction and flat fielding) and aperture photometry were performed using IRAF \footnote{IRAF is distributed by the national optical astronomy observatories, USA.} software. The R-band differential photometry (i.e., variable minus comparison star) was done by using the same comparison star as used by \citet{2007A&A...473..511S}.

\begin{deluxetable}{lccccccccccccccc}
\tablecolumns{4}
\tablewidth{-0pt}
\tabletypesize{\scriptsize}
\setlength{\tabcolsep}{0.05in}
\tablecaption{Periods corresponding to dominant peaks in the power spectra of Paloma obtained from X-ray and optical data.\label{tbl-1}}
\tablehead{
  &\multicolumn{2}{c}{\textbf{X-ray}} & \textbf{Optical}\\
\cline{2-4}
\\    
  &  PN   &MOS   & Photometry
\\
  & (0.3$-$10.0 keV)      &(0.3$-$10.0 keV)}
\startdata
Identification&Period&Period& Period\\
              & (minutes)&(minutes)&(minutes)\\   
\hline\hline
\\
$P$$_\Omega$&156$\pm$1&158$\pm$1& 157.1$\pm$0.2\\
$P$$_\omega$&130$\pm$1&127$\pm$2&136.36$\pm$0.01\\
$P$$_\omega${$^\prime$} &--&--&147.5$\pm$0.1 \\
$P$$_($$_2$$_\omega$$_-$$_\Omega$$_)$& 111$\pm$2&113$\pm$3&--\\
$P$$_2$$_\Omega$&75$\pm$1&73$\pm$1&78.6$\pm$0.1\\
$P$$_3$$_\Omega$&50.6$\pm$0.2&50.5$\pm$0.4& --\\
$P$$_($$_\omega$$_+$$_2$$_\Omega$$_)$&47.5$\pm$0.1&47.5$\pm$0.1& --\\
$P$$_3$$_\omega$&44.3$\pm$0.2& 44.3$\pm$0.4& --\\
$P$$_($$_2$$_\omega$$_+$$_2$$_\Omega$$_)$&36.4$\pm$0.2&36.5$\pm$0.3& --\\
$P$$_($$_2$$_\omega$$_+$$_3$$_\Omega$$_)$&29.7$\pm$0.1&29.6$\pm$0.1& --\\
$P$$_($$_\omega$$_+$$_\Omega$$_)$&--&--& 73.00$\pm$0.04\\
$P$$_{\omega}$$_-$$_\Omega$$_{,}$$_{obs}$  &780$\pm$85&668$\pm$121&1002.5$\pm$1.0\\
\enddata
\end{deluxetable}

\section{Analysis And Results}
\subsection{X-Ray Data}
\subsubsection{Light Curves and Power Spectra}
The background-subtracted X-ray light curves obtained from three EPIC instruments in the soft ($0.3 - 2.0$ keV), hard ($2.0 - 10.0$ keV), and total ($0.3 - 10.0$ keV) energy bands along with the hardness ratio curve are shown in Figure \ref{fig:pnmoslc}. The hardness ratio (HR) is defined as (H$-$S)\small/(H$+$S); where H and S are count rates in hard and soft energy bands, respectively. The binning time of all light curves is 200 s. The 

\begin{deluxetable*}{lcccrcrcccr}
\tablecolumns{7}
\tablewidth{-0pt}
\tabletypesize{\tiny}
\setlength{\tabcolsep}{0.05in}
\tablecaption{Values of X-ray spectral parameters\label{tbl-2} Obtained from the Best-fit models to the data.}
\tablehead{
Models(~$\rightarrow$)     &\colhead{A}&\colhead{ B}&\colhead{C}&\colhead{D}  &\colhead{E}&\colhead{ F} 
\\                 
Parameters($\downarrow$)   &  &   & &  & & }
\startdata
tbabs ($N$$_{\rm H}$$_1$)                     & 4.4 $_{-0.4}^{+0.4}$      & 7.5 $_{-1.6}^{+1.8}$      & 7.9$_{-0.9}^{+0.9}$      & \colhead{2.8 $_{-0.5}^{+0.5}$}     & 3.8 $_{-0.4}^{+0.4}$ &  \colhead{6.6$_{-1.0}^{+1.0}$} \\
pcfabs ($N$$_{\rm H}$$_2$)                    & 5.5$_{-0.3}^{+0.3}$       & 6.0$_{-0.4}^{+0.4}$       & 6.2$_{-0.4}^{+0.4}$      & \colhead{3.0$_{-0.4}^{+0.4}$}      & 4.0$_{-0.4}^{+0.4}$ & \colhead{4.6$_{-0.5}^{+0.6}$}   \\
pcf                                   & 0.66$_{-0.01}^{+0.01}$    & 0.65$_{-0.01}^{+0.01}$    & 0.64$_{-0.01}^{+0.01}$   & \colhead{0.55$_{-0.03}^{+0.03}$}   & 0.59$_{-0.02}^{+0.02}$ & \colhead{0.60$_{-0.02}^{+0.02}$}  \\
reflect ($R$$_{refl}$)                  & \nodata                   & \nodata                   & \nodata                  & \colhead{\nodata}                  & 2.8$_{-0.7}^{+0.7}$   & \colhead{2.4$_{-0.7}^{+0.8}$}    \\
bb ($kT$)                               & \nodata                   & 0.09$_{-0.01}^{+0.01}$    & \nodata                  & \colhead{2.6$_{-0.2}^{+0.3}$}      & \nodata & \colhead{\nodata}               \\ 
$n$ ($\times$10\textsuperscript{-5})    & \nodata                   & 0.7$_{-0.3}^{+0.3}$       & \nodata                  & \colhead{5.0$_{-0.6}^{+0.6}$}      & \nodata      & \colhead{\nodata}       \\
apec ($ kT$$_1$ )                        & \nodata                   & \nodata                   & 0.11$_{-0.01}^{+0.01}$   & \colhead{1.4$_{-0.1}^{+0.2}$}      &  \nodata & \colhead{0.10$_{-0.01}^{+0.02}$} \\ 
$n$$_1$ ($\times$10\textsuperscript{-3})& \nodata                   & \nodata                   & 1.0$_{-0.4}^{+0.5}$      & \colhead{0.07$_{-0.02}^{+0.03}$}   &  \nodata & \colhead{0.3$_{-0.2}^{+0.3}$}    \\
apec ({\it kT}$_2$)                         & 12.6$_{-0.6}^{+0.6}$      & 12.6$_{-0.6}^{+0.5}$      & 12.6$_{-0.6}^{+0.6}$     & \colhead{10.1$_{-0.8}^{+0.8}$}     & 13.1$_{-0.5}^{+1.4}$ & \colhead{13.0$_{-0.5}^{+0.5}$}   \\
$n$$_2$ ($\times$10\textsuperscript{-3})& 4.4$_{-0.1}^{+0.1}$       & 4.4$_{-0.1}^{+0.1}$       & 4.4$_{-0.1}^{+0.1}$      & \colhead{2.5$_{-0.3}^{+0.3}$}      & 3.5$_{-0.2}^{+0.2}$ & \colhead{3.7$_{-0.2}^{+0.2}$}    \\
EW                 & 116$_{-19}^{+18}$         & 113$_{-19}^{+18}$         & 112$_{-18}^{+17}$        & \colhead {92$_{-18}^{+17}$}         & 95$_{-17}^{+18 }$                            &\colhead{97$_{-17}^{+17 }$}\\
$n$$_g$ ($\times$10\textsuperscript{-5})& 1.0$_{-0.2}^{+0.2}$       & 1.0$_{-0.1}^{+0.1}$       & 0.9$_{-0.1}^{+0.1}$      & \colhead{0.8$_{-0.1}^{+0.1}$}      & 0.6$_{-0.1}^{+0.1}$  & \colhead{0.7$_{-0.1}^{+0.1}$}   \\
X-ray flux ($f_X$)                    & 0.610$_{-0.006}^{+0.006}$ & 0.630$_{-0.010}^{+0.011}$ & 0.610$_{-0.006}^{+0.006}$& \colhead{0.630$_{-0.007}^{+0.007}$}& 0.625$_{-0.007}^{+0.007}$ & \colhead{0.640$_{-0.008}^{+0.008}$}  \\
X-ray luminosity ($L_X$)              & 4.18$_{-0.04}^{+0.04}$    & 4.32$_{-0.07}^{+0.08}$    & 4.18$_{-0.04}^{+0.04}$   & \colhead{4.32$_{-0.05}^{+0.05}$}   & 4.30$_{-0.05}^{+0.05}$                        & \colhead{4.40$_{-0.05}^{+0.05}$}                      \\
$\chi_\nu^2$ (dof)                    & 1.16(1782)                & 1.15(1780)                & 1.13(1780)               & \colhead{1.09(1778)}               & 1.13(1781)                                    & \colhead{1.12(1779)} 
\enddata
\tablecomments 
{Where models A$-$tbabs$\times$pcfabs(apec+Gauss), B$-$tbabs$\times$pcfabs(bb+apec+Gauss), C$-$tbabs$\times$pcfabs
(apec+apec+Gauss), D$-$tbabs$\times$pcfabs(bb+apec+apec+Gauss), E$-$tbabs$\times$reflect$\times$pcfabs(apec+Gauss), and F$-$tbabs$\times$reflect$\times$pcfabs(apec+apec+Gauss). $N$$_{\rm H}$$_1$ $=$ Absorption due to galactic hydrogen column in units of 10\textsuperscript{20} cm \textsuperscript{-2}, $N$$_{\rm H}$$_2$ is the partial covering absorber density, i.e., absorption due to partial covering of the X-ray source by the neutral hydrogen column in units of 10\textsuperscript{22} cm \textsuperscript{-2}, pcf is the covering fraction of the partial absorber, bb is the blackbody temperature in keV, $n$ is the normalization constant of the blackbody, $kT$$_1$ and $kT$$_2$ are apec temperatures in units of keV, $n$$_1$ and $n$$_2$ are the normalization constants of apec, EW is equivalent width of the Fe K$\alpha$ line in units of eV, $n$$_g$ is normalization constant of Gaussian, $f_X$ is the X-ray flux derived for 0.3-10.0 keV energy band in units of 10\textsuperscript{-11}erg cm \textsuperscript{-2} s \textsuperscript{-1}, and $L_X$ is the X-ray luminosity calculated in units of 10\textsuperscript{31}erg s \textsuperscript{-1} by assuming a distance of 240 pc. All the errors are with a 90\% confidence interval for a single parameter ($\Delta$ $\chi^2$$=$2.706).} 
\end{deluxetable*}

\noindent
light curves are continuous and show periodic intensity variations. We performed a Fourier Transform (FT) of the {\it XMM-Newton} data using Lomb$-$Scargle periodogram \citep{1976Ap&SS..39..447L, 1982ApJ...263..835S, 1986ApJ...302..757H} as implemented in the {\sc Starlink Period Program}. Figure \ref{fig:powspeclc} show the Lomb$-$Scargle power spectra of EPIC-PN timing data in the total, soft, and hard energy bands. We have also calculated the False Alarm Probability \citep[see][]{1986ApJ...302..757H} in order to check the significance of detected peaks. The horizontal line in Figure \ref{fig:powspeclc} shows a 99\% significance level. The power spectra of MOS and PN data show consistent periods corresponding to the peak frequencies and are given in Table 1. The frequencies in the power spectra as obtained from the PN data correspond to the longest periods of 156 $\pm$ 1 minutes and 130 $\pm$ 1 minutes which appear to be the orbital and spin periods, respectively.  The error on these periods were derived by using the method given in \citet{1986ApJ...302..757H}. Following the above interpretation, the beat period ({\it P}$_{beat}$ = {\it P}$_\Omega$ {\it P}$_\omega$\small/{\it P}$_\Omega$ $-$ {\it P}$_\omega$) was derived to be 780 minutes from the present X-ray data. The beat period also coincides with a sub-harmonic of orbital period 5 $\times$ P$_\Omega$ and spin period 6 $\times$ P$_\omega$. The frequency corresponding to the beat period was also present in the power spectra and marked by $\omega$$-$$\Omega$. The other dominant peaks in the power spectra correspond to periods of {\it P}$_($$_2$$_\omega$$_-$$_\Omega$$_)$ =  111 $\pm$ 2 minutes, {\it P}$_2$$_\Omega$ = 75 $\pm$ 1 minutes, {\it P}$_3$$_\Omega$ = 50.6 $\pm$ 0.2 minutes, {\it P}$_($$_\omega$$_+$$_2$$_\Omega$$_)$ =  47.5 $\pm$ 0.1 minutes, {\it P}$_3$$_\omega$ =  44.3 $\pm$ 0.2 minutes, {\it P}$_($$_2$$_\omega$$_+$$_2$$_\Omega$$_)$ =  36.4 $\pm$ 0.2 minutes, and {\it P}$_($$_2$$_\omega$$_+$$_3$$_\Omega$$_)$ = 29.7 $\pm$ 0.1 minutes. The power spectra in all energy bands show similar peak frequencies. The peak powers corresponding to the periods {\it P}$_($$_\omega$$_-$$_\Omega$$_)$, {\it P}$_\Omega$, {\it P}$_($$_\omega$$_+$$_2$$_\Omega$$_)$, and {\it P}$_($$_2$$_\omega$$_+$$_3$$_\Omega$$_)$ were found to be more in the hard band than those in the soft band. However, powers of all other peaks were found to be more in the soft band compared to that in the hard band. We have folded the PN data using the ephemeris HJD$_0$ = 2450897.35376 as given by \citet{2007A&A...473..511S}. The folded X-ray light curves and HR curves corresponding to the derived orbital and spin periods are shown in Figure \ref{fig:pnmosfoldlc}. The periodic modulation in all energy bands is clearly seen. Orbital and spin period modulations were also seen in HR curve, which is an indication of the dependence of spectral parameters on the orbital and spin phases.

\begin{figure*}
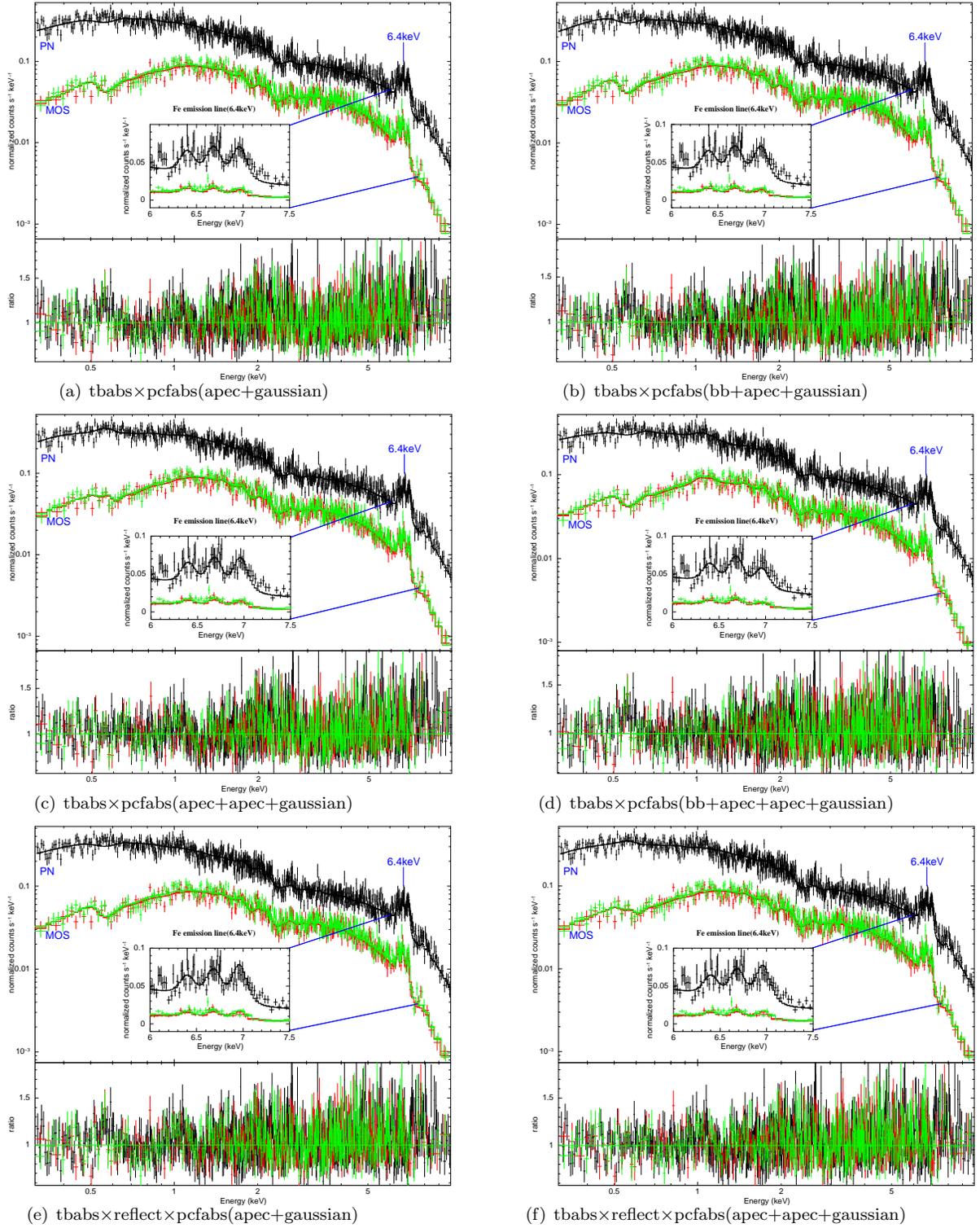

\centering
\subfigure[tbabs$\times$pcfabs(apec+gaussian)]{\includegraphics[width=62mm,angle=-90]{fig4a.ps}}
\subfigure[tbabs$\times$pcfabs(bb+apec+gaussian)]{\includegraphics[width=62mm,angle=-90]{fig4b.ps}}
\subfigure[tbabs$\times$pcfabs(apec+apec+gaussian)]{\includegraphics[width=62mm,angle=-90]{fig4c.ps}}
\subfigure[tbabs$\times$pcfabs(bb+apec+apec+gaussian)]{\includegraphics[width=62mm,angle=-90]{fig4d.ps}}
\subfigure[tbabs$\times$reflect$\times$pcfabs(apec+gaussian)]{\includegraphics[width=62mm,angle=-90]{fig4e.ps}}
\subfigure[tbabs$\times$reflect$\times$pcfabs(apec+apec+gaussian)]{\includegraphics[width=62mm,angle=-90]{fig4f.ps}}
\caption{EPIC-PN (top) and MOS (bottom) spectra of Paloma fitted simultaneously with the six composite models. The zoomed spectra and the model fit around the Fe-K$\alpha$ region are shown in the inset. The bottom panel shows the $\chi^2$ contribution of data points for the best-fitting model in terms of ratio.}
\label{fig:pnmosspec}
\end{figure*}

\subsubsection{X-ray Spectral Analysis}
 The background-subtracted X-ray spectra of Paloma are shown in Figure \ref{fig:pnmosspec}, where strong broadband continuum along with the strong emission line features at 6.4 keV are clearly seen. The inset in Figure \ref{fig:pnmosspec} shows an enlarged view of the $6.0 - 7.5$ keV energy range, which contains the Fe K$\alpha$ fluorescent line at 6.4 keV and 6.7, 6.95 keV lines from ionized Fe {\sc xxv} and Fe {\sc xxvi}, respectively. Spectral fits were performed on the PN and MOS data using XSPEC version$-12.9.0$ \citep{1996ASPC..101...17A, 2001ASPC..238..415D} with various models\small/combinations of models. These models were blackbody (bb), bremsstrahlung (bremss), astrophysical plasma emission code \citep[apec;][]{2001ApJ...556L..91S}, and Gaussian. A common absorption component (tbabs) was used in all models/combinations of models. A single bremsstrahlung or blackbody temperature model with tbabs component produced an unacceptable fit with reduced $\chi^2_\nu$ $\geq$ 3.0. The apec single temperature plasma emission model with the tbabs component also yielded an unacceptably high value of $\chi^2_\nu$ of 3.09. Though the apec plasma emission model can account for the continuum and several emission lines, it is not able to account for the strong Fe line at 6.4 keV. Therefore, an additional Gaussian line at 6.4 keV was added to the single component of apec resulting in an improved $\chi^2_\nu$ of 1.5 (dof 1790). All the above combinations of models gave unacceptably high values of $\chi^2$. Therefore, we fitted all the above models/combinations of models with partially covering absorption model (pcfabs), which gave a better fit but still with an unacceptable $\chi^2_\nu$. However, the model tbabs$\times$pcfabs(apec+Gauss) [model A] gave an acceptable value of reduced $\chi^2_\nu$ of 1.16 (dof 1782), but this model was unable to explain the excess around 0.5 keV, which is clearly seen in the ratio plot of Figure \ref{fig:pnmosspec} (a) indicating the presence of low-temperature component. We therefore, added a blackbody component along with the single apec temperature as tbabs$\times$pcfabs(bb+apec+Gauss) [model  B], but there was no improvement in the $\chi^2_\nu$ and the excess near 0.5 keV was still present on the spectra (see Figure \ref{fig:pnmosspec} (b)). The use of two apec components along with a Gaussian, modeled as tbabs$\times$pcfabs(apec+apec+Gauss) [model  C] with solar photospheric abundances from \citet{1989GeCoA..53..197A}, slightly improved the ($\chi^2_\nu$ = 1.13, dof = 1780), and provides a better fit to the spectra near 0.5 keV (see Figure \ref{fig:pnmosspec} (c)). In order to obtain the best fit, we have also fitted the spectra by adding a blackbody model as tbabs$\times$pcfabs(bb+apec+apec+Gauss) [model D], which improves the $\chi^2_\nu$ to 1.09, (dof = 1778) but the excess near 0.5 keV is still present (see Figure \ref{fig:pnmosspec} (d)). In all four models, we obtained the equivalent width (EW) of the Fe K$\alpha$ emission line more than $\sim$ 100 eV (see Table 2). To account for the derived large EW, the column density of cold matter should be as high as 10\textsuperscript{23} cm \textsuperscript{-2} \citep[see][]{1985SSRv...40..317I}, while this value is found a factor of $\sim$ 10 lower for the partial absorber in models A, B, C, and D. This suggests the presence of the contribution of a Compton reflected continuum from the WD, which is expected to go along with the fluorescent iron line \citep{1991A&A...247...25M, 1992ApJ...395..275D, 1999ApJS..120..277E}. Therefore, to take into account a reflected continuum of the plasma emission due to the Fe K$\alpha$ at 6.4 keV, we have also used the reflection model, which is a convolution type of model describing the reflectivity of a neutral material \citep{1995MNRAS.273..837M}. Including this new model component with a combination of other models as tbabs$\times$reflect$\times$pcfabs(apec+Gauss) [model E], the model parameters remain the same as in model A and the excess was also seen near 0.5 keV (see Figure \ref{fig:pnmosspec} (e)). Finally, we adopt model F for the spectral fitting, which is tbabs$\times$reflect$\times$pcfabs(apec+apec+Gaussian). This model improves  the value of $\chi^2_\nu$ to 1.12 (dof=1779) and also fit the spectra near 0.5 keV (see Figure \ref{fig:pnmosspec} (f)). In the above combinations of models keeping the line width parameter free in the Gaussian model does not fit the Gaussian line well and parameter value pegged at the upper limit. Therefore, to fix the value of the line width in the Gaussian, we changed the line width with the values 0.005, 0.01, 0.02, 0.05, 0.1, 0.15, and 0.2 keV and found that the values of line widths $\textgreater$ 0.02 keV begin to merge the line with the continuum, while when line widths $\textless$ 0.02 keV are narrower than observed and in both cases there is no change in the value of $\chi^2$ and spectral parameters. The best fit was thus obtained when we have fixed the width of the Gaussian line at 0.02 keV. The spectral parameters derived from the simultaneous fitting to both EPIC-PN and MOS spectra using model A, B, C, D, E, and F are given in Table 2. The error bars quoted here are with 90\% confidence limit for a single variable parameter. The best-fit value of the partial absorber hydrogen column density ({\it N}$_{\rm H}$$_2$) was 4.6$_{-0.5}^{+0.6}$ $\times$ 10\textsuperscript{22} cm \textsuperscript{-2} with a covering fraction (pcf) of 60 $\pm$ 2 percent and apec temperatures ($kT$$_1$ and $kT$$_2$) of 0.10$_{-0.01}^{+0.02}$ and 13.0$_{-0.5}^{+0.5}$ keV. The galactic hydrogen column density ({\it N}$_{\rm H}$$_1$) corrected flux ({\it f}$_X$) was calculated by using the `cflux' model. The X-ray luminosity of Paloma was calculated by assuming a distance of 240 pc \citep{2007A&A...473..511S} in the energy band $0.3-10.0$ keV and is given in Table 2.

\begin{figure}
\centering
\hspace {-1.0cm}
\includegraphics[width=65mm,angle=-90]{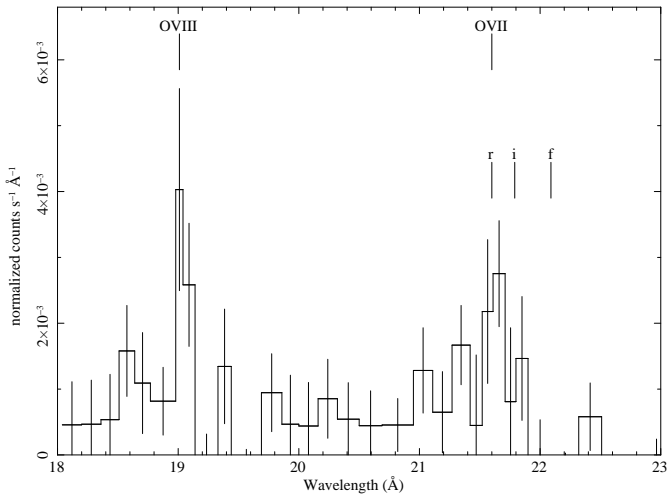}
\caption{RGS spectra near the O {\sc viii} and O {\sc vii} line.}
\label{fig:rgsspec}
\end{figure}

Spectral data from RGS were of poorer quality compared to PN and MOS data and are shown in Figure \ref{fig:rgsspec}. The O {\sc viii} Ly$\alpha$ at 19.1 {\AA} and the O {\sc vii} He-like triplet at 21.9 {\AA} are seen in the RGS spectra. The detected O {\sc vii} line was likely to be a blend of the resonant (r), intercombination (i), and forbidden (f) lines, but with this noisy data spectra fits with any value of model (Gaussian) parameter. Therefore, finally, we have not attempted the spectral model fitting on RGS data.

\begin{deluxetable*}{lcccrcrcccr}
\centering
\tablecolumns{9}
\tablewidth{-0pt}
\tabletypesize{\scriptsize}
\setlength{\tabcolsep}{0.05in}
\tablecaption{\leftskip0mm{The best fit X-ray spectral parameters derived from the analysis of PN and MOS data at different orbital and spin phases\label{tbl-3}.}}
\tablehead{
\\
Model        & Parameters & \multicolumn{5}{c}{Orbital Phase}\\
\cline{3-7}\\ 
             &    &  0.0$-$0.2      &   0.2$-$0.4   &  0.4$-$0.6     & 0.6$-$0.8 & 0.8$-$1.0}
\startdata                                                                        
pcfabs &$N$$_{\rm H}$$_2$($\times$10\textsuperscript{22} cm \textsuperscript{-2})&6.2$_{-1.2}^{+1.2}$    &10.0$_{-4.3}^{+6.5}$   & 4.8$_{-0.8}^{+0.9}$   &3.8$_{-0.6}^{+0.7}$&3.0$_{-0.5}^{+0.6}$\\

       & pcf                                                                   & 0.53$_{-0.05}^{+0.02}$&0.38$_{-0.13}^{+0.04}$ & 0.61$_{-0.04}^{+0.04}$& 0.63$_{-0.03}^{+0.03}$&0.56$_{-0.03}^{+0.03}$\\

reflect&$R$$_{refl}$& 0.00005$_{-0.00002}^{+0.00002}$    &0.0034$_{-0.0008}^{+0.0008}$   & 1.74$_{-1.35}^{+1.52}$   &5.80$_{-1.60}^{+1.70}$ & 3.7$_{-1.2}^{+1.3}$ \\

apec& $n$$_1$($\times$10\textsuperscript{-3}) & 0.15$_{-0.07}^{+0.07}$ & 0.70$_{-0.04}^{+0.04}$ & 0.36$_{-0.11}^{+0.11}$ & 0.42$_{-0.12}^{+0.13}$ & 0.58$_{-0.12}^{+0.12}$\\

apec& $n$$_2$($\times$10\textsuperscript{-3}) & 2.31$_{-0.02}^{+0.05}$ & 1.27$_{-0.22}^{+0.03}$ & 3.98$_{-0.40}^{+0.40}$ & 4.49$_{-0.40}^{+0.45}$ & 4.89$_{-0.36}^{+0.36}$\\

Gaussian (6.4 keV)& $n$$_g$($\times$10\textsuperscript{-5})         & 0.4$_{-0.2}^{+0.2}$ &0.2$_{-0.1}^{+0.1}$  &0.7$_{-0.3}^{+0.3}$  & 0.8$_{-0.3}^{+0.3}$&0.9$_{-0.3}^{+0.3}$\\

EW & Fe K$\alpha$ (eV)  & 85$_{-57}^{+69}$ & 40$_{-27}^{+27}$ & 99$_{-44}^{+41}$ & 91$_{-33}^{+32}$ & 94$_{-34}^{+31}$\\

X-ray flux (0.3$-$10.0 keV)       & $f_X$($\times$10\textsuperscript{-11}erg cm \textsuperscript{-2} s \textsuperscript{-1})    & 0.36$_{-0.01}^{+0.01}$ & 0.21$_{-0.01}^{+0.01}$ & 0.66$_{-0.02}^{+0.02}$ & 0.91$_{-0.02}^{+0.02}$ & 0.96$_{-0.02}^{+0.02}$\\

X-ray luminosity (0.3$-$10.0 keV) & $L_X$($\times$10\textsuperscript{31}erg s \textsuperscript{-1})                             &2.47$_{-0.07}^{+0.07}$ & 1.44$_{-0.07}^{+0.07}$ & 4.53$_{-0.13}^{+0.13}$ & 6.24$_{-0.13}^{+0.13}$ & 6.59$_{-0.13}^{+0.13}$\\

                              & $\chi_\nu^2$ (dof) &1.13(358)&1.28(245)&1.05(555)&1.10(675)&0.91(752)\\
\hline\\
\centering 
Model&Parameters&\multicolumn{5}{c}{Spin Phase}\\
\cline{3-7}\\ 
     &    &  0.0$-$0.2      &   0.2$-$0.4   &  0.4$-$0.6     & 0.6$-$0.8 & 0.8$-$1.0\\
\hline\\                                                                          
pcfabs&$N$$_{\rm H}$$_2$($\times$10\textsuperscript{22} cm \textsuperscript{-2})&4.6$_{-1.0}^{+1.3}$&3.4$_{-0.6}^{+0.7}$&4.3$_{-1.0}^{+1.0}$&4.8$_{-1.0}^{+1.2}$&6.8$_{-1.5}^{+1.7}$\\

& pcf                                                                 & 0.54$_{-0.06}^{+0.05}$& 0.58$_{-0.04}^{+0.03}$&0.56$_{-0.05}^{+0.05}$&0.60$_{-0.05}^{+0.05}$&0.64$_{-0.06}^{+0.03}$\\                                                   

reflect&$R$$_{refl}$& 2.01$_{-1.24}^{+1.40}$ & 2.80$_{-1.80}^{+2.05}$ & 0.83$_{-0.01}^{+0.03}$  & 3.27$_{-1.73}^{+1.96}$   &3.91$_{-2.00}^{+2.32}$   \\

apec& $n$$_1$($\times$10\textsuperscript{-3}) &0.28$_{-0.09}^{+0.09}$&0.53$_{-0.12}^{+0.13}$& 0.15$_{-0.10}^{+0.10}$&0.37$_{-0.10}^{+0.12}$&0.24$_{-0.10}^{+0.11}$\\       

apec& $n$$_2$($\times$10\textsuperscript{-3})    &3.18$_{-0.40}^{+0.47}$& 4.37$_{-0.36}^{+0.41}$& 3.50$_{-0.40}^{+0.46}$&3.15$_{-0.40}^{+0.48}$&3.27$_{-0.50}^{+0.35}$\\    

Gaussian (6.4 keV)& $n$$_g$($\times$10\textsuperscript{-5})   & 0.3$_{-0.2}^{+0.2}$&1.0$_{-0.3}^{+0.4}$&0.8$_{-0.2}^{+0.3}$&0.7$_{-0.3}^{+0.3}$&0.4$_{-0.3}^{+0.4}$\\    
                
EW & Fe K$\alpha$ (eV)  & 50$_{-44}^{+44}$ & 83$_{-41}^{+40}$ & 51$_{-42}^{+43}$ & 103$_{-48}^{+47}$ & 132$_{-48}^{+58 }$\\

X-ray flux (0.3$-$10.0 keV)& $f_X$($\times$10\textsuperscript{-11}erg cm \textsuperscript{-2} s \textsuperscript{-1})&0.58$_{-0.01}^{+0.01}$&0.80$_{-0.02}^{+0.02}$&0.64$_{-0.01}^{+0.01}$&0.58$_{-0.01}^{+0.01}$&0.48$_{-0.01}^{+0.01}$\\            
                                                                                        X-ray luminosity (0.3$-$10.0 keV) & $L_X$($\times$10\textsuperscript{31}erg s \textsuperscript{-1})&3.98$_{-0.07}^{+0.07}$&5.49$_{-0.13}^{+0.13}$&4.39$_{-0.07}^{+0.07}$&3.98$_{-0.07}^{+0.07}$&3.29$_{-0.07}^{+0.07}$\\

                              & $\chi_\nu^2$ (dof)&0.90(458)&1.04(591)&0.93(491)&1.21(445)&1.05(384)\\

\enddata

\tablenotetext{~}{{\it Note:} Quoted error bars are with a 90\% confidence limit for a single parameter.  }
\end{deluxetable*}

\begin{figure*}
\centering
\includegraphics[width=140mm,angle=0]{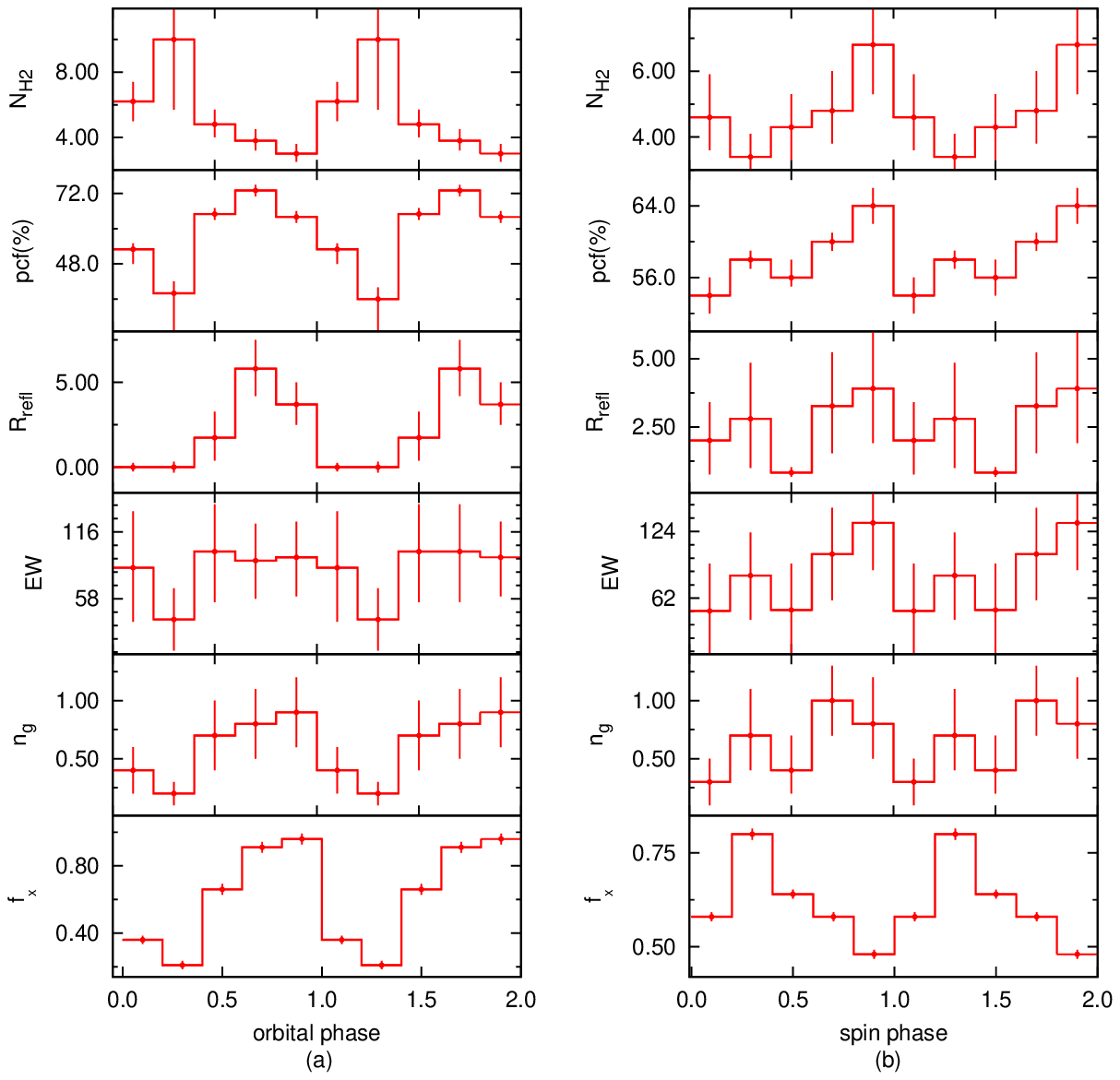}
\caption{Variation of the best-fit spectral parameters as a function of orbital and spin phases. In both Figures (a) and (b), top to bottom panels show the variation of column densities ($N$$_{\rm H}$$_2$) due to partial covering of the X-ray source by neutral hydrogen, partial covering fraction (pcf), relative normalization of the reflection component ($R$$_{refl}$), equivalent width (EW) of fluorescence iron line, the normalization of Gaussian ($n$$_g$) component, and flux ($f_X$), where ($N$$_{\rm H}$$_2$) in units of 10\textsuperscript{22} cm \textsuperscript{-2}, EW in units of eV, $n$$_g$ in units of 10\textsuperscript{-5} photons cm$^{-2}$ s$^{-1}$, and flux in units of 10\textsuperscript{-11} erg cm \textsuperscript{-2} s \textsuperscript{-1}. Error bars are plotted with a 90\% confidence limit for a single parameter.}
\label{fig:specparm}
\end{figure*}

\subsubsection{Orbital and Spin-phase-resolved X-ray Spectroscopy}
In order to trace the dependence of X-ray spectral parameters on orbital and spin phases, we performed orbital and spin-phase-resolved spectroscopy. The orbital and spin-phase spectra from EPIC-PN and MOS data were extracted for all five equal phase intervals of 0.2 from phase 0 to 1. Extracted spectra from both the detectors were simultaneously fitted for each interval with the best-fit model F. Initially, we allowed the apec temperatures to vary during the fit in all the five (orbital and spin) phase intervals, but that adjusted the value of other parameters such that the spectral fitting gave unacceptably high value of $\chi^2$. Therefore, the {\it N}$_{\rm H}$$_1$ and temperatures of apec component ({\it kT}$_1$ and {\it kT}$_2$) were fixed at the values obtained from average spectral fitting. The free parameters were {\it N}$_{\rm H}$$_2$, pcf, relative normalization of the reflection component ({\it R}$_{refl}$), normalizations of apec ({\it n}$_1$ and {\it n}$_2$), and Gauss ($n_g$) models. We also derived the EW from the spectral fit in each orbital and spin-phase segment. Individual parameters obtained from the orbital and spin-phase-resolved spectral analysis are given in Table 3. The variations in {\it N}$_{\rm H}$$_2$, pcf, {\it R}$_{refl}$, EW, {\it n}$_g$, and unabsorbed flux {\it f}$_X$ are shown in Figure \ref{fig:specparm}. We find an anti-correlation between {\it f}$_X$ and {\it N}$_{\rm H}$$_2$ in both (orbital and spin) phase spectral analyses, while pcf and {\it f}$_X$ are correlated with the orbital phase and anti-correlated with the spin phase. The variations of the {\it R}$_{refl}$ and EW of the fluorescent line increases at the spin minimum phase. The normalization of the Fe K$\alpha$ emission line is also dependent on the spin and the orbital phases, i.e., the variations in line flux are correlated with changes in {\it f}$_X$.

\subsection{Optical Photometry}

\begin{figure*}
\centering
\includegraphics[width=115mm,angle=-90]{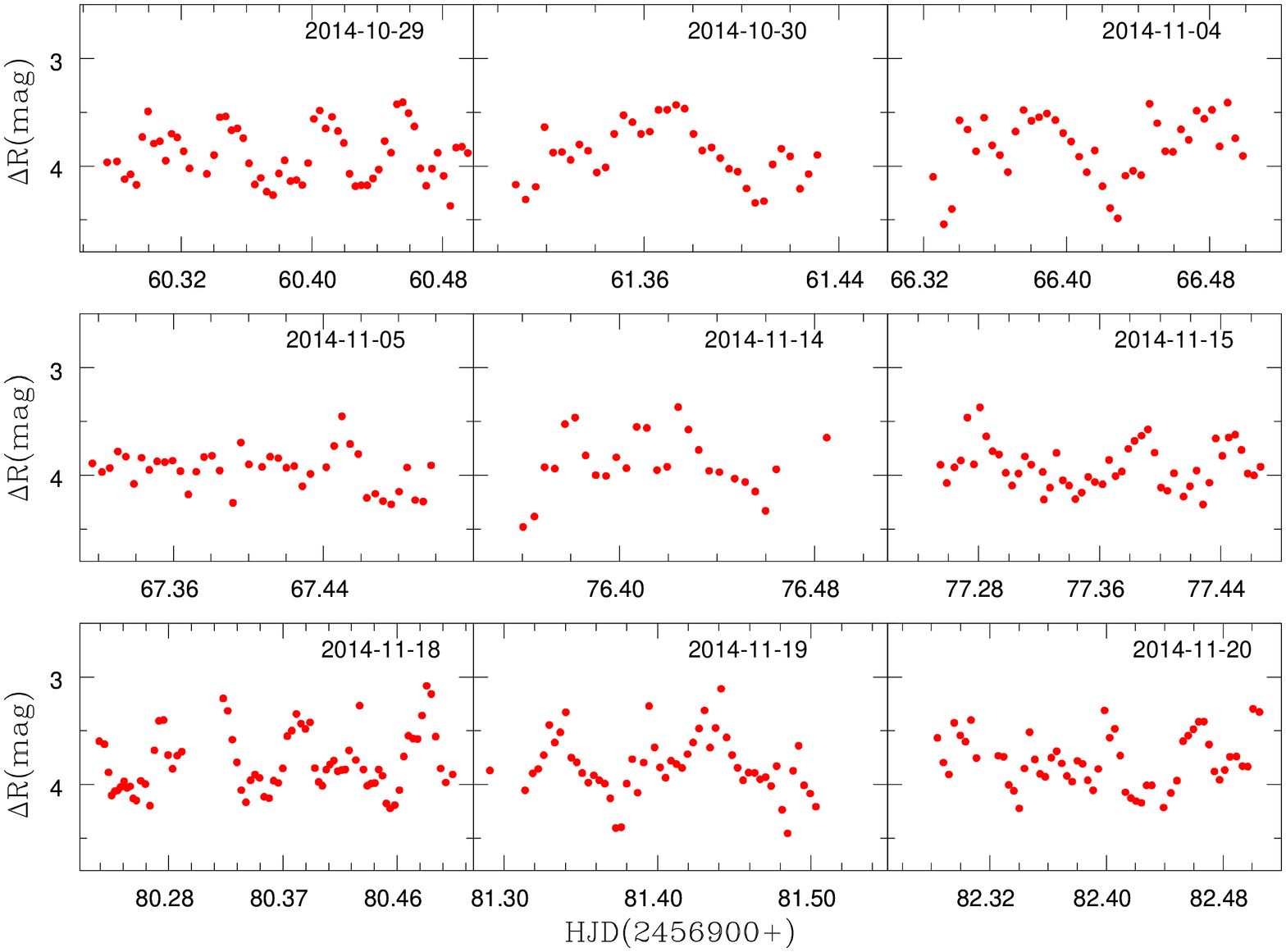}
\caption{R-band light curve of Paloma. The observed dates are mentioned at the top of each panel.}
\label{fig:photlc}
\end{figure*}

\begin{figure}
\centering
\vspace{0.5cm}
\includegraphics[width=90mm,angle=0]{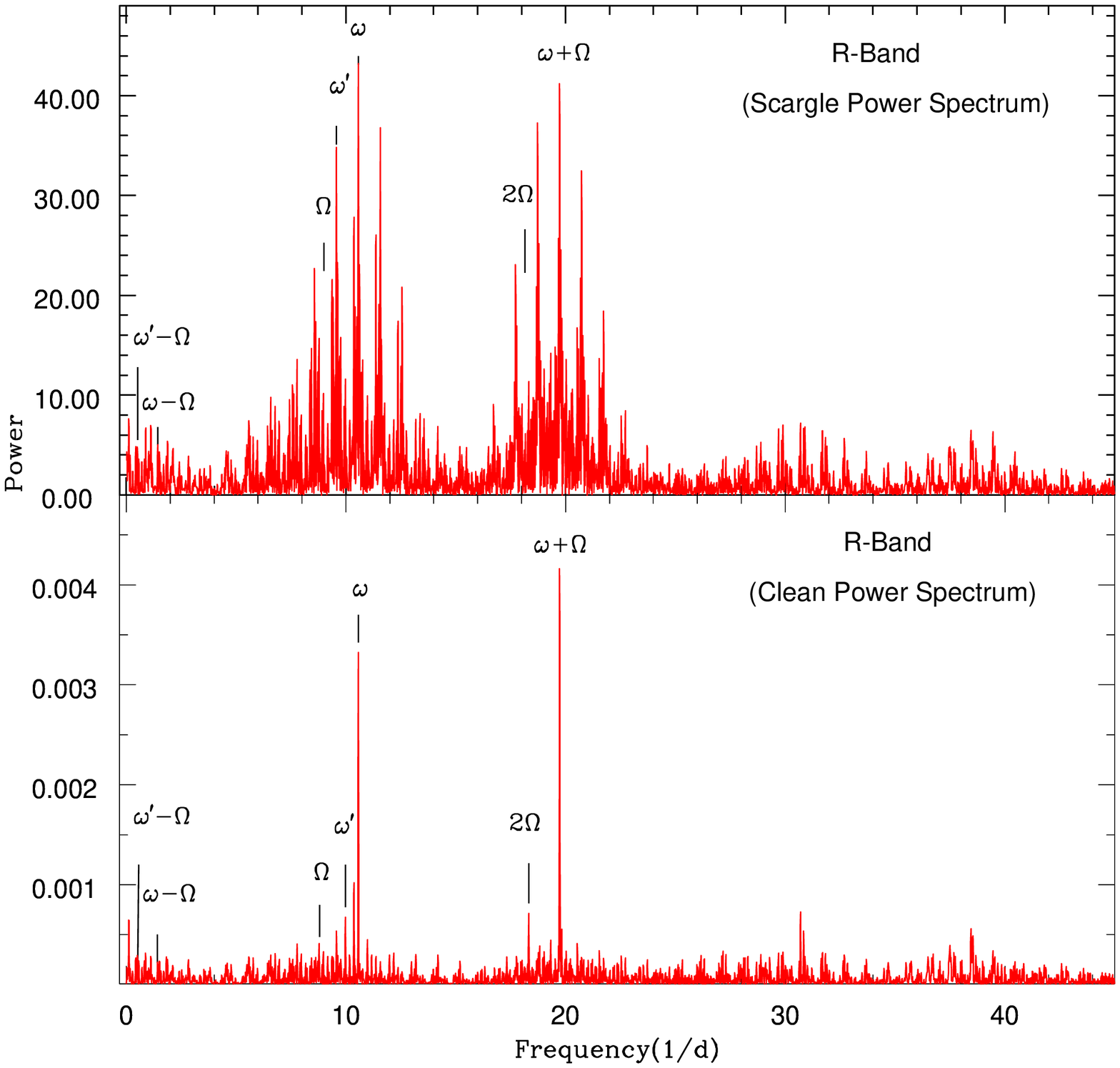}
\caption{Power density spectrum of Paloma, computed from the combined photometric data. The top panel shows power spectra obtained using the Lomb$-$Scargle Periodogram, while the bottom panel shows the CLEANed power density spectrum.}
\label{fig:optpowlc}
\end{figure}

\begin{figure*}
\centering
\subfigure[]{\includegraphics[width=110mm,angle=-90]{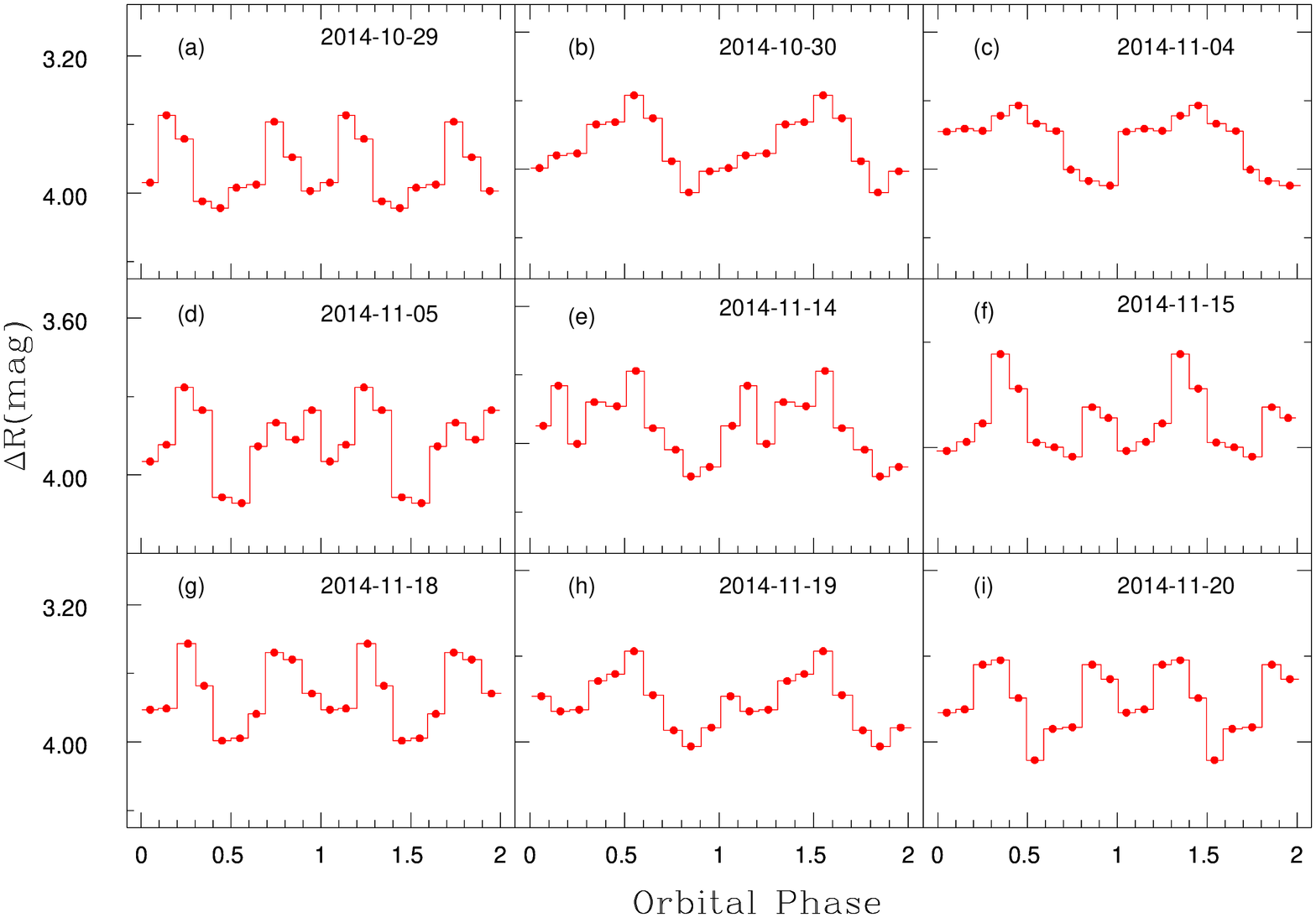}}
\subfigure[]{\includegraphics[width=110mm,angle=-90]{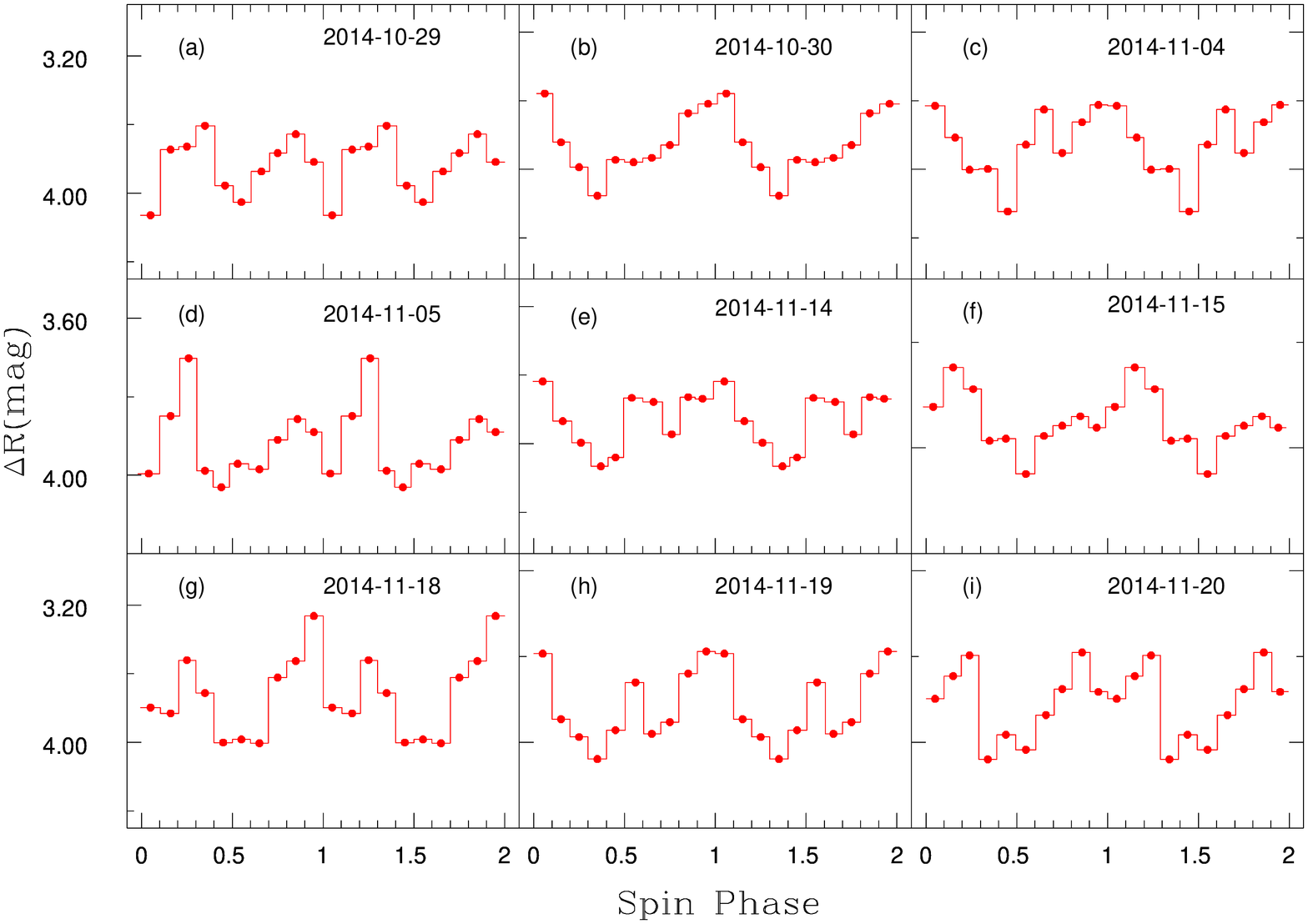}}
\caption{Orbital and spin-phase-folded light curves of Paloma in the R-Band on different dates of observations.}
\label{fig:orbspinfoldlc}
\end{figure*}

Figure \ref{fig:photlc} shows the R-band light curve of Paloma for different nights. It shows multiscale time variability and strong flickering with optical pulsations in the light curve. The light curve appeared to show either a single hump or double-humped periodic variation. The Lomb$-$Scargle power spectra of optical R-band data is shown in the top panel of Figure \ref{fig:optpowlc}, which is noisier compared to the power spectrum from X-ray data. Therefore, the CLEAN algorithm \citep{1987AJ.....93..968R} was applied to the data, which basically deconvolves the spectral window from the discrete Fourier power spectrum (or dirty spectrum) and produces a CLEAN spectrum. The CLEANed power spectra presented here were obtained after 100 iterations of the CLEAN procedure with a loop gain of 0.1. We found six significant peaks corresponding to periods given in Table 1. These periods correspond to the frequency components at ${\omega}$$-$$\Omega$, $\Omega$, $\omega${$^\prime$}, $\omega$, 2$\Omega$, and $\omega$+$\Omega$; where $\omega$ and $_\omega${$^\prime$} are two spin periods. The error in the period calculation was set by the finite resolution of the power spectrum and was determined as {\it P}$^2$\small/2{\it t}$_{max}$, where {\it P} is the period and {\it t}$_{max}$ is the duration of the observations \citep[see][]{1987AJ.....93..968R}. The photometric data were folded using the periods {\it P}$_\Omega$ and {\it P}$_\omega$ and are shown in Figures \ref{fig:orbspinfoldlc} (a) and (b), respectively. On October 29, and November 18 and 20, clear double-humped structure was seen in orbital as well as spin-phase-folded light curves, but it disappeared on the next night, October 30. On the other hand, November 5, 14, 15, and 19 orbital phase-folded light curves seem to show a double-humped structure, while in the spin-phase-folded light curve the second peak is less prominent. On October 30 and November 4 modulations with orbital phases appear to be maximum around phases ($\sim$ $0.5 - 0.6$) and ($\sim$ $0.4 - 0.5$), respectively, while these variations are interrupted by a broad, eclipse-like feature centered around the spin phases ($\sim$ $0.3 - 0.4$) and ($\sim$ $0.4 - 0.5$) on both nights, respectively, with an amplitude of 0.8 mag.  
 
\section{Discussion}
We have carried out analyses of X-ray and optical photometric data of an MCV Paloma. Using the X-ray data from {\it XMM-Newton}, we derive 10 significant periods in Paloma, including two fundamental periods of 156 $\pm$ 1 minutes and 130 $\pm$ 1 minutes. However, using our optical data, we derive seven significant periods in which {\it P}$_\Omega$ and {\it P}$_\omega$ are consistent with $\sim$1.5 $\sigma$ level to that derived from X-ray data. The peaks corresponding to the frequencies $\Omega$, $\omega$, $2\Omega$, and $\omega+\Omega$ are well matched with those derived by \citet{2007A&A...473..511S}. The orbital modulation in the X-ray light curves is generally explained by the absorption of X-rays emitted near the WD surface by material fixed in the orbital frame, but a strong signal at the orbital period in the X-ray power spectra of Paloma indicates that an explicit orbital modulation is most likely due to the attenuation introduced by the impact site of the accretion stream with the accretion disk or magnetosphere. On the other hand, X-ray modulation at the spin period can arise due to the self-occultation of emission regions by the WD or due to photoelectric absorption and electron scattering in the accretion curtain \citep{1996MNRAS.280..937N}. The occurrence of the spin peak indicates that the symmetry of the two pole caps is broken and the amount of asymmetry depends on whether the signal corresponding to the rotation of the asymmetrically placed pole is strong or weak, i.e., the degree of asymmetry is proportional to the power of the spin peak \citep{1992MNRAS.255...83W}. Along with the spin frequency, the presence of significant peaks in the X-ray power spectrum at $\omega$-$\Omega$, $\Omega$, $2\omega-\Omega$, $2\Omega$, $3\Omega$, $\omega+2\Omega$, $3\omega$, $2\omega+2\Omega$, and $2\omega+3\Omega$ also predict that Paloma is an asymmetric system in which both the emission poles are not diametrically opposite and have slightly different emission properties \citep[see][]{1996MNRAS.280..937N}. Furthermore, the presence of significant power at spin, beat and  $2\omega-\Omega$ frequencies suggest that Paloma is a disk-less accreting system \citep[see][]{1992MNRAS.255...83W,1996MNRAS.280..937N}. Additionally, an increase in power of the beat or $2\omega-\Omega$ components accompanied by the decrease in power at the spin component as the X-ray energy increases is a characteristic of the disk-less accretion. This is clearly visible in the X-ray power density spectra of Paloma, which is further attributed to the absorption of low-energy photons by the accretion column itself. Other than the orbital, spin, and beat periods, photometric observations of Paloma show peaks at frequencies $2\Omega$ and $\omega+\Omega$. Such similar positive and negative sideband frequencies corresponding to the spin and orbital periods are also observed from photometric observations in nearly synchronous IP V697 Sco \citep{2002PASP..114.1222W}, suggesting that Paloma may have a close resemblance to V697 Sco.

MCVs occupy a wide range of parameter space in the {\it P}$_\omega$ $-$ {\it P}$_\Omega$ plane in which the evolution of these systems can be drawn from the degree of synchronization with the orbital period. Paloma is one of the few MCVs with an orbital period lying in the period gap ($\sim$ $2 - 3$ hr) and discovered as one of the nearly synchronous IPs \citep[see][]{2004ASPC..315..216N}. The asynchronocity (1 $-$ {\it P}$_\omega$\small /{\it P}$_\Omega$) of Paloma was found to be $\sim$ 16.6 \% from the X-ray and optical data, which is slightly away from the line of synchronization. Furthermore, the values of {\it P}$_($$_\omega$, $_\omega${$^\prime$}$_)$\small /{\it P}$_\Omega$ ($\sim$ 0.8) obtained from X-ray and optical analyses of Paloma satisfy the synchronization condition derived by \citet{2004ASPC..315..216N} i.e., {\it P}$_\omega$/{\it P}$_\Omega$ $\textgreater$ 0.6. The observed asynchronocity in Paloma suggests that it might be representative of a transition object currently in the process of attaining synchronism and evolving into polars.

In the X-ray power spectrum of Paloma, we have also found a specific signal at the beat frequency along with a long WD spin period ({\it P}$_\omega$ $\sim$ 130 minutes), and X-ray modulations show  a single-peaked pulse profile as seen in some IPs FO Aqr, TX Col, BG CMi, AO Psc, V1223 Sgr, and RX J1712.6-2414. This indicates that accretion occurs in a stream-fed scenario where the WD has  a relatively strong magnetic field \citep[see][]{1999A&A...347..203N}. Unlike the single-peaked X-ray modulations, such modulations were not detected in the optical domain of Paloma along with a long WD spin period. The amplitude of the optical modulations varies greatly from night to night, which is a common behavior for the IPs \citep[see, for example,][]{1996MNRAS.282..739W}. The disappearance of the double hump from one night of observation to another indicates that the beat period of Paloma is much shorter than the other near-synchronous polars.

 We see that the amplitude of the rotational pulsation is independent of energy. This could be due to the occultation of X-ray emitting regions as WD rotates. However, there are many IPs where the amplitude of rotational pulsation increases with decreasing energy, which is consistent with photoelectric absorption in the accretion flow.  

We have noticed the excess in the X-ray spectra of Paloma near 0.5 keV, which is well explained by low-temperature apec component. In order to account this excess, we have derived the unabsorbed soft ({\it F}$_s$) and hard ({\it F}$_h$) X-ray fluxes of 2.1$_{-0.2}^{+0.2}$ $\times$10\textsuperscript{-13} and 7.7$_{-0.1}^{+0.1}$ $\times$10\textsuperscript{-12} erg cm\textsuperscript{-2} s\textsuperscript{-1} in the energy band of 0.3 - 10.0 keV from the low-{\it kT} apec model and high-{\it kT} apec model, respectively. The softness ratio is then calculated by using the formula {\it F}$_s$/4{\it F$_h$} and the value is obtained as 0.0068, which closely matches with the softness ratio of IP V2400 Oph \citep{2007ApJ...663.1277E}. The soft-X-ray excess is mostly prominent in the X-ray spectra of polars; however, the growing number of IPs (e.g., PQ Gem, NY Lup, GK Per, V405 Aur, and V2400 Oph) also shows the soft X-ray component. In many systems, this soft-X-ray excess was explained by the black-body component with a temperature less than 100 eV  \citep[e.g.,][]{1992MNRAS.258..749M, 1994A&A...291..171H, 2004A&A...415.1009D, 2007ApJ...663.1277E, 2008A&A...489.1243A, 2009A&A...501.1047A}. However, in an IP IGR J17195$-$4100 the soft-X-ray excess was well fitted by a low-temperature apec component in X-ray spectra \citep{2012MNRAS.427..458G}. \citet{1982A&A...114L...4K} suggest that the soft-X-ray excess usually seen in polars due to blobby accretion, i.e., dense blobs of matter penetrate into the WD photosphere and the energy is thermalized to a blackbody, but the main reason some IPs show soft-X-ray excess is simply that the heated region near the accretion footprints is not hidden by the accretion curtains while, in other IPs, the accretion footprint is hidden by the accretion curtains, the difference being the result of the system inclination and the magnetic colatitude \citep[see][]{2007ApJ...663.1277E}.

The variations in the absorption component ({\it N}$_{\rm H}$$_2$) are found to be minimum when the X-ray fluxes are maximum for both the spin and the orbital cycles. The self-absorption can cause such variation in an IP with the accretion curtain. The X-ray flux will be minimum due to the maximum absorption when the curtain is maximally in the line of sight, and the X-ray flux will be maximum when either the curtain moves away from the line of sight or when it thins down. The high value of the X-ray flux with the spin phase requires a small value of the covering fraction of the source by an absorber. This implies the least hindered view toward the accreting pole at this spin phase. The variations in the relative normalization of the reflection component are consistent with the EW of the iron emission line and their contribution is maximum at the rotational minimum phase. The correlated variations of pcf, EW, and {\it R}$_{refl}$ with spin phase (in Figure \ref{fig:specparm}) indicate that these variations support the accretion curtain scenario. The pcf, EW, and {\it R}$_{refl}$ are found to be maximum at the minimum value of flux at spin phase. This could be due to the fact that the projected accretion area on the WD surface is larger and pointed toward the observer. The variations in the iron K$\alpha$ emission line flux closely resemble the variations in the X-ray flux in the 0.3$-$10.0 keV energy band, indicating that in the region where the line is produced there is no change in the accretion geometry, and are also visible at all times during the binary and rotation motion. The correlation of line flux with the average X-ray intensity may also indicate that the derived orbital and spin periods are true periods.

\section{Summary}
To summarize, we find the following characteristics of Paloma, which are well explained by present optical and X-ray data.
\begin{enumerate}[nolistsep]
\item Orbital and spin periods determined from the timing analysis of X-ray and optical data indicate that Paloma lies slightly away from the line of synchronization and it could be a transition object between polars and IPs.
\item The X-ray power spectra obtained in different energy bands and the presence of significant peaks at spin, beat, and $2\omega-\Omega$ frequencies suggest that Paloma is a disk-less system.
\item The X-ray pulse is energy independent, which indicates that the spin pulsations could be originating due to the occultation of the emitting region by rotation of WD.
\item  The soft-X-ray excess observed in the X-ray spectra of Paloma could be due to the accretion curtains, which are unable to hide the accretion footprints due to the highly inclined magnetic axis of the system.
\item  A marginal detection of O {\sc viii} and O {\sc vii} lines in the RGS spectra indicates the presence of the cooler optically thin emitting region.
\end{enumerate}

\acknowledgments
We are grateful to the referee for useful comments and suggestions that improved the manuscript considerably. This research is based on observations obtained by {\it XMM-Newton}, an ESA science mission with instruments and contributions directly funded by ESA Member States and the USA (NASA).

{\it Facility: \facility {XMM-Newton}}   

\bibliography{ref}
\bibliographystyle{apj}

\end{document}